\documentclass[sigconf, authorversion]{acmart}

\AtBeginDocument{%
  \providecommand\BibTeX{{%
    \normalfont B\kern-0.5em{\scshape i\kern-0.25em b}\kern-0.8em\TeX}}}

\copyrightyear{2023} 
\acmYear{2023} 
\setcopyright{acmlicensed}
\acmConference[AISec '23]{Proceedings of the 16th ACM Workshop on Artificial Intelligence and Security}{November 30, 2023}{Copenhagen, Denmark}
\acmBooktitle{Proceedings of the 16th ACM Workshop on Artificial Intelligence and Security (AISec '23), November 30, 2023, Copenhagen, Denmark}
\acmPrice{15.00}
\acmDOI{10.1145/3605764.3623920}
\acmISBN{979-8-4007-0260-0/23/11}

\usepackage{amsmath,amsfonts}
\usepackage{algorithmic}
\usepackage{graphicx}
\usepackage{xcolor}
\usepackage{xspace}
\usepackage{hyperref}
\usepackage{listings}
\usepackage{enumitem}
\usepackage{multirow}
\usepackage{booktabs}
\usepackage{array}
\usepackage{subcaption}
\usepackage{pifont}
\usepackage{fancyvrb}
\usepackage{optidef}
\usepackage{multicol}

\usepackage[ruled,linesnumbered]{algorithm2e}

\usepackage[acronym]{glossaries}

\makeatletter
\newcommand\inlinecode[1]
  {\lstinline[basicstyle=\ttfamily]{#1}}
\makeatother

\makeatletter
\newcommand\inlinehtml[1]
  {\lstinline[language=html]{#1}}
\makeatother

\graphicspath{{figs/}}

\hypersetup{%
  colorlinks=true,
  linkcolor=black,
  citecolor=black,
  urlcolor=black
}

\newacronym{ML}{ML}{machine learning}
\newacronym{JS}{JS}{JavaScript}
\newacronym{RL}{RL}{Reinforcement Learning}
\newacronym{AI}{AI}{Artificial Intelligence}
\newacronym{TNR}{TNR}{True Negative Rate}
\newacronym{FPR}{FPR}{False Positive Rate}
\newacronym{TPR}{TPR}{True Positive Rate}
\newacronym{ROC}{ROC}{Receiver Operating Characteristic}
\newacronym{DR}{DR}{Detection Rate}
\newacronym{MLSEC}{MLSEC}{Machine Learning Security Evasion Competition}

\graphicspath{{figures/}}

\newcolumntype{C}[1]{>{\centering\let\newline\\\arraybackslash\hspace{0pt}}m{#1}}

\makeatletter
\AtBeginDocument{\let\c@listing\c@lstlisting}
\makeatother

\begin{document}

\title{Raze to the Ground: Query-Efficient Adversarial HTML Attacks on Machine-Learning Phishing Webpage Detectors}

\author{Biagio Montaruli}
\email{biagio.montaruli@sap.com}
\affiliation{%
  \institution{SAP Security Research \& EURECOM}
  \city{Mougins}
  \country{France}
}

\author{Luca Demetrio}
\email{luca.demetrio@unige.it}
\affiliation{%
  \institution{University of Genova \& Pluribus One}
  \city{Genova}
  \country{Italy}
}

\author{Maura Pintor}
\email{maura.pintor@unica.it}
\affiliation{%
  \institution{University of Cagliari \& Pluribus One}
  \city{Cagliari}
  \country{Italy}
}

\author{Luca Compagna}
\email{luca.compagna@sap.com}
\affiliation{%
  \institution{SAP Security Research}
  \city{Mougins}
  \country{France}
}

\author{Davide Balzarotti}
\email{davide.balzarotti@eurecom.fr}
\affiliation{%
  \institution{EURECOM}
  \city{Biot}
  \country{France}
}

\author{Battista Biggio}
\email{battista.biggio@unica.it}
\affiliation{%
  \institution{University of Cagliari \& Pluribus One}
  \city{Cagliari}
  \country{Italy}
}

\renewcommand{\shortauthors}{Biagio Montaruli, et al.}

\newcommand{\sota}{state-of-the-art\xspace}
\newcommand{\diff}[2]{\frac{\partial #1}{\partial #2}}
\newcommand{\vct}[1]{\ensuremath{\boldsymbol{#1}}}
\newcommand{\mat}[1]{\ensuremath{\mathbf{#1}}}
\newcommand{\set}[1]{\ensuremath{\mathcal{#1}}}
\newcommand{\con}[1]{\ensuremath{\mathsf{#1}}}
\newcommand{\tsum}{\ensuremath{\textstyle \sum}}
\newcommand{\T}{\ensuremath{\top}}
\newcommand{\mycomment}[1]{\footnote{\textcolor{red}{#1}}}
\newcommand{\ind}[1]{\ensuremath{\mathbbm 1_{#1}}}
\newcommand{\argmax}{\operatornamewithlimits{\arg\,\max}}
\newcommand{\erf}{\text{erf}}
\newcommand{\sign}{\text{sign}}
\newcommand{\argmin}{\operatornamewithlimits{\arg\,\min}}
\newcommand{\bmat}[1]{\begin{bmatrix}#1\end{bmatrix}}
\newcommand{\myparagraph}[1]{\noindent \textbf{#1}}
\newcommand{\myparagraphlb}[1]{\noindent \newline \textbf{#1}}
\newcommand{\mysubparagraph}[1]{\noindent \underline{\textit{#1}}}
\newcommand{\ie}{i.e.\xspace}
\newcommand{\eg}{e.g.\xspace}
\newcommand{\etc}{etc.\xspace}
\newcommand{\aka}{a.k.a.\xspace}
\newcommand{\wrt}{w.r.t.\xspace}
\newcommand{\etal}{\emph{et al.}\xspace}

\newcommand{\myceil}[1]{\left \lceil #1 \right \rceil }

\newcommand\smamath[1]{{\small $#1$}}
\newcommand\res[2]{\small{$#1$}\tiny{\text{$\pm #2$}}}
\newcommand\bestres[2]{\small{$\mathbf{#1}$}\tiny{\text{$\mathbf{\pm #2}$}}}
\newcommand\smacal[1]{{\small $\mathcal{#1}$}}
\newcolumntype{?}{@{\hskip\tabcolsep\vrule width 1pt\hskip\tabcolsep}}

\newcommand{\wafamole}{WAF-A-MoLE\xspace}

\makeatletter
\newenvironment{mcases}[1][l]
 {\let\@ifnextchar\new@ifnextchar
  \left\lbrace
  \def\arraystretch{1.2}%
  \array{@{}l@{\quad}#1@{}}}
 {\endarray\right.}
\makeatother

\begin{abstract}
Machine-learning phishing webpage detectors (ML-PWD) have been shown to suffer from adversarial manipulations of the HTML code of the input webpage.
Nevertheless, the attacks recently proposed have demonstrated limited effectiveness due to their lack of optimizing the usage of the adopted manipulations, and they focus solely on specific elements of the HTML code.
In this work, we overcome these limitations by first designing a novel set of fine-grained manipulations which allow to modify the HTML code of the input phishing webpage without compromising its maliciousness and visual appearance, i.e., the manipulations are functionality- and rendering-preserving by design.
We then select which manipulations should be applied to bypass the target detector by a query-efficient black-box optimization algorithm.
Our experiments show that our attacks are able to \emph{raze to the ground} the performance of current state-of-the-art ML-PWD using just 30 queries, thus overcoming the weaker attacks developed in previous work, and enabling a much fairer robustness evaluation of ML-PWD.
\end{abstract}



\maketitle

\section{Introduction}\label{sec:introduction}

\begin{figure}[h]
    \centering
    \includegraphics[width=\columnwidth]{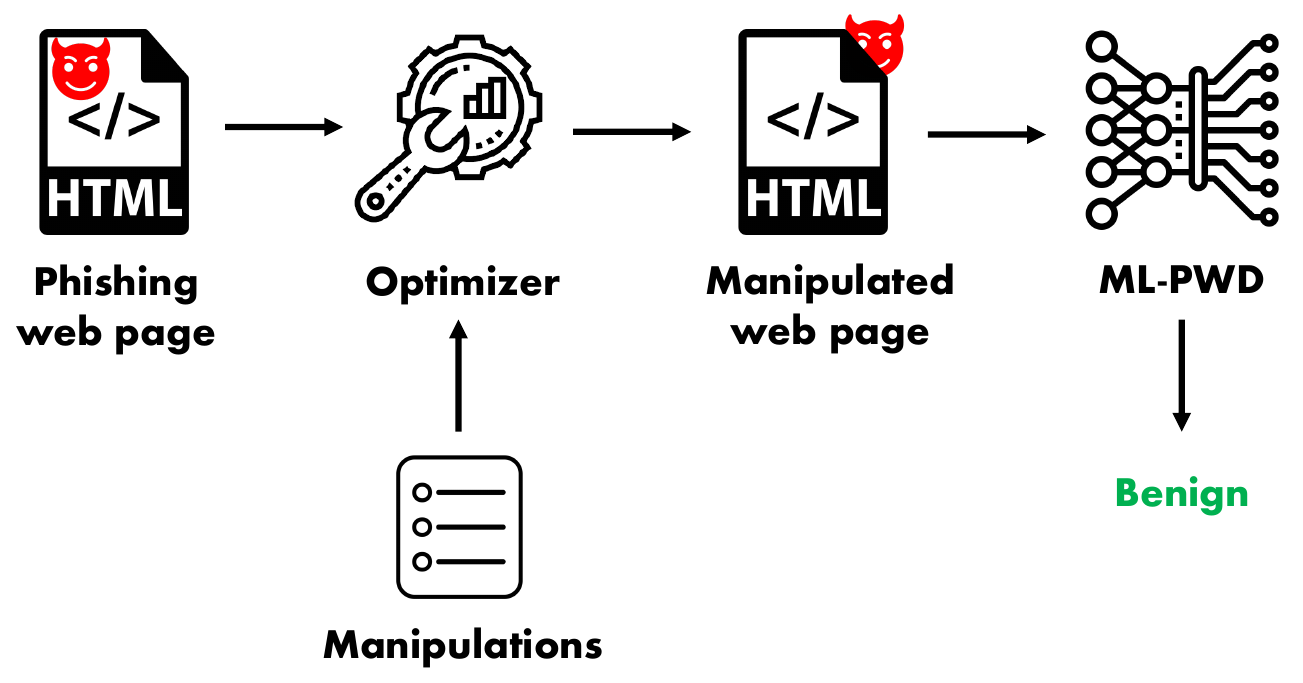}
    \caption{Overview of our work: we propose a novel set of adversarial manipulations that are functionality- and rendering-preserving by design, and a query-efficient black-box optimizer to generate HTML adversarial attacks that are able to \emph{raze to the ground} \sota machine-learning phishing webpage detectors (ML-PWD).}
\label{fig:system_overview}
\end{figure}

Over the past years, we witnessed a significant increase in the number of phishing attacks \cite{kaspersky_phishing_2022, vade_phishing_2022, proofpoint2023phish}, thereby emphasizing that this remains a significant form of cybercrime.
Among all the different types of phishing, this work focuses on the detection of phishing \emph{webpages}, which are typically created by an attacker to steal sensitive information such as login credentials~\cite{Apruzzese:SpacePhish}.
To counter this open problem, in addition to the use of blocklists \cite{prakash2010phishnet,oest2020phishtime} that have been demonstrated easy to bypass by \emph{adaptive} attackers \cite{Tian:TrackingElitePhishing}, novel approaches based on machine-learning \cite{Tian:TrackingElitePhishing, tang2021survey, niakanlahiji2018phishmon, Jain2018TowardsDO, Sharma2020:FeatSelectionPhishing, Hannousse:Benchmark_Phishing, Corona:Deltaphish} have been proposed in recent years to enhance the detection capabilities of phishing detection systems.
However, phishing webpage detectors based on machine-learning (\ie, ML-PWD, using the same acronym of Apruzzese \etal \cite{Apruzzese:SpacePhish}) have been shown to be vulnerable to adversarial attacks \cite{Apruzzese:SpacePhish, Corona:Deltaphish, liang2016cracking, Bac:PWDGAN, abdelnabi2020visualphishnet, Bahnsen:2018DeepPhishS, aleroud2020bypassing, Gressel:ModelAgnosticAdvAttack}, both in the \emph{problem space}, which is the input space of HTML pages, and the \emph{feature space}, which is the space where webpages are represented as feature vectors~\cite{arp2022dos, Apruzzese:SpacePhish}.
In problem-space attacks, the attacker directly manipulates the URL \cite{Bahnsen:2018DeepPhishS, Bac:PWDGAN}, the HTML code \cite{li2014towards, liang2016cracking} or the visual representation \cite{abdelnabi2020visualphishnet} of the phishing webpage with physically realizable manipulations \cite{arp2022dos}, while feature-space attacks only manipulate the abstract feature representation of input samples.
To this extent, \emph{SpacePhish} \cite{Apruzzese:SpacePhish} represents one of the most recent and comprehensive studies about adversarial attacks against ML-PWD both in the problem and feature spaces.
Indeed, its authors provide a well-validated benchmark of \sota ML-PWD.

However, their work is characterized by two major limitations.
First, investigating how the adversarial robustness changes if the attacker is able to optimize the adversarial attacks by querying the target ML-PWD is an important open point of their work.
Second, they use a limited set of \emph{cheap} adversarial manipulations \ie, manipulations that do not require any knowledge about the structure of phishing webpages such as the insertion of internal links and URL-shortening \cite{Apruzzese:SpacePhish, Apruzzese:SpacePhish_Supp}, which result in weak or, in some cases, even useless attacks.
Indeed, the reported results show that, for some evaluated ML-PWD, such \emph{cheap} attacks (indicated as ${WA}^r$ in their paper) cause the manipulated phishing webpage to appear even \textit{more malicious} to the ML-PWD.

To address the aforementioned limitations, we propose a novel methodology for generating optimized and query-efficient HTML adversarial attacks (see \autoref{fig:system_overview}).
Specifically, we first perform a thorough security analysis of the HTML features used in \emph{SpacePhish} (Sect.~\ref{sec:background}), which are widely adopted in the literature~\cite{Hannousse:Benchmark_Phishing, Jain2018TowardsDO, Mohammad2014IntelligentRP, Sharma2020:FeatSelectionPhishing}, to understand how they can be evaded.
Then, in addition to the HTML manipulations proposed in \emph{SpacePhish}, we design a novel set of 14 manipulations that maintain the original functionality \cite{demetrio2021functionality} and rendering \cite{Gao:MLSEC} while manipulating the HTML code of phishing webpages (Sect.~\ref{sec:manipulations}).
On the basis of these manipulations, we formulate a query-efficient black-box optimization algorithm (Sect.~\ref{sec:mutation-algo}) that generates optimized adversarial phishing webpages in the problem space.

Finally, we validate our approach through an extensive experimental analysis (Sect.~\ref{sec:experiments}), showing that our novel adversarial attacks are able to completely evade the ML-PWD evaluated in \emph{SpacePhish} using just 30 queries.
To foster reproducible results, we share the source code of our work\footnote{\url{https://github.com/advmlphish/raze_to_the_ground_aisec23}}.

To summarize, we provide the following three contributions:
\begin{itemize}
    \item We conduct a comprehensive security analysis of the HTML features used in \emph{SpacePhish} and, on top of it, we devise a novel set of adversarial manipulations that are functionality- and rendering-preserving by design, with the goal to evade all the analyzed features.
    \item We propose a black-box optimizer inspired by mutation-based fuzzing \cite{fuzzingbook2023:MutationFuzzer}, which allows to craft optimized HTML adversarial attacks using the proposed manipulations;
    \item We empirically show that our methodology allows to \emph{raze to the ground} the detection capabilities of current \sota ML-PWD using very few queries.
\end{itemize}
We conclude the paper by discussing the open points of our work, along with promising future research directions (Sect.~\ref{sec:conclusions}).
\section{Background}\label{sec:background}
In this section, we first give an overview of the basic structure of webpages and then we describe the HTML features adopted in \emph{SpacePhish} \cite{Apruzzese:SpacePhish}.

\subsection{Webpage Structure}\label{sec:webpage_overview}
Webpages are generally described using the HTML language~\cite{html_spec}.
They have a basic structure that consists of a tree hierarchy represented by the HTML Document Object Model (DOM) tree~\cite{html_dom}, which is made of multiple HTML elements corresponding to the DOM nodes.
Each HTML element is represented through (i) a single tag or a pair of (start and end) tags and (ii) some content that includes text or other nested HTML elements.
Moreover, HTML elements can have attributes consisting of name-value pairs to provide additional information about the element.
Although the HTML specification includes many types of elements, a typical webpage (see \autoref{code:webpage_example}) includes the head (lines \ref{code_webpage:head_start}-\ref{code_webpage:head_end}) and the body (lines \ref{code_webpage:body_start}-\ref{code_webpage:body_end}) represented by the \lstinline[language=html]{<head>} and \lstinline[language=html]{<body>} element, respectively.
The head is used to set the webpage title through the \lstinline[language=html]{<title>} element (line \ref{code_webpage:title}) and optionally to define the visual appearance of some embedded HTML elements through the \lstinline[language=html]{<style>} element (lines \ref{code_webpage:style_start}-\ref{code_webpage:style_end}).
The body, instead, includes the main content of the webpage, \ie, all the HTML elements that are generally displayed by a web browser.
For instance, the example webpage includes a login form (lines \ref{code_webpage:form_start}-\ref{code_webpage:form_end}), defined via a \lstinline[language=html]{<form>} element, which consists of two \lstinline[language=html]{<input>} elements used to collect the username (line \ref{code_webpage:input_username}) and password (line \ref{code_webpage:input_pass}) from the user.

\begin{lstlisting}[
   language=HTML,
   float=htb,
   basicstyle=\fontsize{7.5}{8.5}\selectfont\ttfamily,
   commentstyle=\color{gray},
   showspaces=false,
   showstringspaces=false,
   breakindent=1em,
   breaklines=true,
   numbers=left,
   xleftmargin=1.6em,
   keepspaces=true,
   frame=single,
   caption={Example of a webpage.},
   captionpos=b,
   escapechar=|,
   label={code:webpage_example}
]
<!DOCTYPE html>
<html>
<head>  |\label{code_webpage:head_start}|
<title>Website title</title>  |\label{code_webpage:title}|
<style> |\label{code_webpage:style_start}|
  h1 {color: red;}
</style>  |\label{code_webpage:style_end}|
</head>  |\label{code_webpage:head_end}|
<body>  |\label{code_webpage:body_start}|
  <h1>Welcome to the website</h1>  |\label{code_webpage:heading}|
  <form action="login.php", method="get">  |\label{code_webpage:form_start}|
    <label for="pwd">Enter your username: </label>
    <input type="text" name="username" required>  |\label{code_webpage:input_username}|
    <label for="pwd">Enter your password: </label>
    <input type="password" name="pass" required>  |\label{code_webpage:input_pass}|
  </form>  |\label{code_webpage:form_end}|
</body>  |\label{code_webpage:body_end}|
</html>
\end{lstlisting}

\subsection{HTML Feature Analysis}\label{sec:html_features}
In the following we analyze in details the features used in \emph{SpacePhish} to better understand how they work and thus, how to evade them.
This is an missing point in \emph{SpacePhish}.
Indeed, its authors only provide a brief description of some of them (5 out 22) in the related supplementary document \cite{Apruzzese:SpacePhish_Supp}, and do not carefully analyze how they can bypassed using problem-space manipulations.
We also remark that such features are also widely used in other papers~\cite{Hannousse:Benchmark_Phishing, Jain2018TowardsDO, Mohammad2014IntelligentRP, Sharma2020:FeatSelectionPhishing}, and some of them also in competitions about machine learning security such as the \gls*{MLSEC}\footnote{\url{https://www.robustintelligence.com/blog-posts/ml-security-evasion-competition-2022}}~\cite{Apruzzese:SpacePhish_Supp, Gao:MLSEC}.

\myparagraph{HTML\_freqDom.}
This feature analyzes the number of internal (n\_int) and external (n\_ext) HTML elements in the webpage.
An element is internal if includes a link that shares the same domain as the webpage URL; otherwise, it is external.
Then, if n\_ext is 0 or n\_int $\geq$ n\_ext, this feature is set to -1 (the webpage is likely benign); else, it is set to +1 (the webpage is likely phishing).
This feature analyzes the following types of HTML elements: anchors (\lstinline[language=html]{<a>}), images (\lstinline[language=html]{<img>}), links (\lstinline[language=html]{<link>}) and videos (\textbf{\lstinline[language=html]{<video>}}).

\myparagraph{HTML\_objectRatio.}
This feature represents the ratio between the number of external HTML elements, n\_ext, and the total one, n\_tot = n\_ext + n\_int, where n\_int represents the number of internal HTML elements.
The ratio is compared against two thresholds: the suspicious ($0.15$) and phishing ($0.30$) thresholds.
If the ratio is lower than the suspicious threshold, the value of the feature is -1 (the webpage is likely benign).
Otherwise, if the ratio is in between the two thresholds, the webpage is assumed suspicious and the feature is set to 0.
Finally, if the ratio is greater than the phishing threshold, the feature is set to +1 (the webpage is likely phishing).
The HTML elements considered by this feature are the same as HTML\_freqDom.

\myparagraph{HTML\_metaScripts.}
This feature is similar to HTML\_objectRatio, but it applies to script (\lstinline[language=html]{<script>}), meta (\lstinline[language=html]{<meta>}) and link (\lstinline[language=html]{<link>}) elements.
This feature adopts two different values for the thresholds. Specifically, the suspicious and phishing thresholds are set to $0.52$ and $0.61$, respectively.
Moreover, if the ratio is greater than $0.61$, the feature is set to +1 (the webpage is likely phishing); if the ratio is less than $0.52$, the feature is -1 (the webpage is likely benign); otherwise, it is set to 0 (the webpage is assumed suspicious).

\myparagraph{HTML\_commPage.}
This feature analyzes the number of internal (n\_int) and external (n\_ext) elements, and is initialized using the following formula:
\begin{equation*}
\text{HTML\_commPage} = \frac{max(\text{n\_ext}, \text{n\_int})}{\text{n\_ext + n\_int}}
\end{equation*}
This feature takes into account the same HTML elements analyzed by both HTML\_objectRatio and HTML\_metaScripts.

\myparagraph{HTML\_commPageFoot.}
This feature works as HTML\_commPage except that it focuses on the HTML elements included in the footer (\lstinline[language=html]{<footer>}) rather than the body of the webpage.

\myparagraph{HTML\_SFH.}
This feature computes the ratio of suspicious forms as the number of suspicious forms divided by the total number of forms.
The ratio is compared against two thresholds: susp\_thr, which is set to 0.5 and is used to decide if a webpage is suspicious, and phish\_thr, which is set to 0.75 and allows to determine whether a webpage is phishing.
According to its implementation, a form is considered suspicious if one of the following conditions is satisfied: it includes an external link (specified through the \lstinline[language=html]{action} attribute), the \lstinline[language=html]{action} attribute is set to "\lstinline{about:blank}" (\ie, it points to a new blank webpage) or when it is set to an empty string (\ie, \lstinline[language=html]{<form action="">}).
In particular, if the ratio is lower than the suspicious threshold, the feature is set to -1 (the webpage is likely benign).
Else, if the ratio is greater than the phishing threshold, then the feature is initialized to +1 (the webpage is likely phishing).
Otherwise, \ie, the ratio is between the two thresholds, this feature is set to 0 (the webpage is considered suspicious).

\myparagraph{HTML\_popUP.}
This feature checks whether the webpage displays a pop-up window that prompts the user for some input, such as credentials in case of phishing webpages.
A pop-up window can be commonly introduced by using the \lstinline[basicstyle=\ttfamily]{prompt()} or \lstinline[basicstyle=\ttfamily]{window.open()} \gls*{JS} functions.
Specifically, this feature looks for the names of such functions and if finds the former, it is set to 1 (the webpage is likely phishing); while it is set to 0 (the webpage is likely suspicious) if finds the latter.
Otherwise, its value is -1 (the webpage is likely benign).

\myparagraph{HTML\_rightClick.}
This feature inspects the source code of the webpage to determine if a context menu has been disabled, which is the equivalent of disabling the mouse right-click.
In particular, it checks the following patterns to disable a context menu: if the \lstinline[basicstyle=\ttfamily]{preventDefault()} method of the HTML DOM is present in the webpage or if there is at least one HTML element with the \lstinline[basicstyle=\ttfamily]{oncontextmenu} attribute set to \lstinline[basicstyle=\ttfamily]{"return false"}.
Hence, this feature is set to +1 (the webpage is likely phishing) if it finds at least one disabled context menu, and to -1 (the webpage is likely benign) otherwise.

\myparagraph{HTML\_domCopyright.}
This feature analyzes if the webpage contains a copyright notice with the copyright symbol (\textcopyright).
If not, the webpage is considered suspicious and its value is set to 0.
Otherwise, if the copyright notice contains the website domain name, the feature is set to -1 (the webpage is likely benign).
Else (\ie, no webpage domain in the copyright notice) it is set to +1 (the webpage is likely phishing).

\myparagraph{HTML\_nullLnkWeb.}
This feature computes the frequency of suspicious anchors contained in a website as the number of suspicious anchors divided by the total number of anchors.
An anchor is considered suspicious if it contains one of the following useless links: "\lstinline[basicstyle=\ttfamily]{#}", "\lstinline[basicstyle=\ttfamily]{#content}", "\lstinline[basicstyle=\ttfamily]{#skip}" and "\lstinline[basicstyle=\ttfamily]{JavaScript ::void(0)}"; or if it is an internal link.

\myparagraph{HTML\_nullLnkFooter.}
This feature works in the same way as HTML\_nullLnkWeb, but it computes the frequency of suspicious anchors included in the footer rather than the body.

\myparagraph{HTML\_brokenLnk.}
This feature computes the ratio of external elements with broken links (\ie, links that point to an unreachable website) against the total number of external ones included in the webpage.
This feature analyzes the same HTML elements considered by both HTML\_objectRatio and HTML\_metaScripts.

\myparagraph{HTML\_loginForm.}
This feature is set to +1 (the webpage is likely phishing) if the webpage contains one or more forms with a useless internal link or an external one; otherwise, it is set to -1 (the webpage is likely benign).
An internal link is useless if it is equal to one of the following: \lstinline[basicstyle=\ttfamily]{""} (empty string), \lstinline[basicstyle=\ttfamily]{#}, \lstinline[basicstyle=\ttfamily]{#nothing}, \lstinline[basicstyle=\ttfamily]{#null}, \lstinline[basicstyle=\ttfamily]{#void}, \lstinline[basicstyle=\ttfamily]{#doesnotexist}, \lstinline[basicstyle=\ttfamily]{#whatever},
\lstinline[basicstyle=\ttfamily]{javascript}, \lstinline[basicstyle=\ttfamily]{javascript::;}, \lstinline[basicstyle=\ttfamily]{javascript::void(0)}, \lstinline[basicstyle=\ttfamily]{javascript::void(0);}.

\myparagraph{HTML\_hiddenDiv.}
This feature checks if there are content division elements, \aka div (\lstinline[language=html]{<div>}), which are hidden by setting the \lstinline{style} attribute to "\lstinline{visibility:hidden}" or "\lstinline{display:none}".
If so, this feature is set to +1 (the webpage is likely phishing), else to -1 (the webpage is likely benign).

\myparagraph{HTML\_hiddenButton.}
This feature is set to +1 (the webpage is likely phishing) if there is at least one button (\lstinline[language=html]{<button>}) element disabled by setting the \lstinline{style} attribute to "\lstinline{disabled}". Otherwise, the webpage is considered benign and this feature is set to -1.

\myparagraph{HTML\_hiddenInput.}
This feature is set to +1 (the webpage is likely phishing) if there is at least one input element that is disabled (\ie, \lstinline[breaklines=true, language=html]{<input disabled>}) or hidden (\ie, \lstinline[breaklines=true, language=html]{<input type="hidden">}). Otherwise, this feature is set to -1 (the webpage is likely benign).

\myparagraph{HTML\_URLBrand.}
This feature analyzes the title (\lstinline[language=html]{<title>}) of the webpage to check whether it contains the website's domain name.
If so, the webpage is considered benign and this feature is set to -1.
Otherwise, it is initialized to +1 (the webpage is likely phishing).
Moreover, if the title is not found, this feature is set to 0 (the webpage is suspicious).

\myparagraph{HTML\_iFrame.}
This feature targets inline frame elements, \aka iframe (\lstinline[language=html]{<iframe>}), usually used to embed a webpage within another one, by checking patterns commonly used for hiding an iframe, such as
\lstinline[language=html,showspaces=false,showstringspaces=false,breaklines=true]{<iframe style="display: none">} and \lstinline[language=html, showspaces=false, showstringspaces=false, breaklines=true]{<iframe style="visibility: hidden">}.
If any of these patterns are found, the feature is set to +1 (the webpage is likely phishing), else to -1 (the webpage is likely benign).

\myparagraph{HTML\_favicon.}
This feature checks if the \emph{favicon} (\ie, an icon associated with a particular website) is loaded from an external source.
If so, it is set to +1 (the webpage is likely phishing), while it is set to -1 (the webpage is likely benign) if the favicon is internal.
Moreover, if no favicon is included in the webpage, it is considered suspicious and this feature is set to 0.
To check the presence of the favicon, this feature looks for link elements including either \lstinline[basicstyle=\ttfamily]{rel="shortcut icon"} or \lstinline[basicstyle=\ttfamily]{rel="icon"} attributes.

\myparagraph{HTML\_statBar.}
This feature inspects the webpage to check whether it changes the text of the status bar at the bottom of the browser window by looking for the presence of \lstinline[basicstyle=\ttfamily]{window.status} in the HTML code.
If so, this feature is set to +1 (the webpage is likely phishing); else the value of the feature is -1 (the webpage is likely benign).

\myparagraph{HTML\_css.}
This feature checks whether the webpage uses an external CSS style sheet, \ie, if the style sheet is imported from an external web location using a link element as in the following example:
\lstinline[breaklines=true, language=html]{<link rel="stylesheet" href="mystyle.css">}.
If so, this feature is set to +1 (the webpage is likely phishing), else to -1 (the webpage is likely benign).

\myparagraph{HTML\_anchors.}
This feature computes the ratio of suspicious anchors included in the webpage and compares it against two thresholds: suspicious ($0.32$) and phishing ($0.505$).
Then, if there are no anchors in the webpage or the ratio is lower than the suspicious threshold, then this feature is set to -1 (the webpage is likely benign).
Else, if the ratio is higher than the phishing threshold, it is set to +1 (the webpage is likely phishing).
Otherwise, \ie, if the ratio is between the two thresholds, its value is 0 (the webpage is considered suspicious).
An anchor is assumed suspicious if contains an external link or if it includes an internal link belonging to the same list of patterns checked by the HTML\_nullLnkWeb feature, \ie, \lstinline[basicstyle=\ttfamily]{#}, \lstinline[basicstyle=\ttfamily]{#skip}, \lstinline[basicstyle=\ttfamily]{#content} and \lstinline[basicstyle=\ttfamily]{JavaScript ::void(0)}.
\section{Threat model}\label{sec:threat_model}
In this section, we first formalize the threat model used in our work, and then we compare it to the one proposed in \emph{SpacePhish} \cite{Apruzzese:SpacePhish}.

\subsection{Formalization}
We describe the threat model according to the following four criteria widely used in the adversarial machine learning literature~\cite{Biggio:Wild}.

\myparagraph{Goal.}
The goal of the adversary consists in causing an integrity violation by evading a target machine-learning phishing detector at test time through adversarial phishing webpages generated in the problem space.
In other words, the adversary aims to manipulate the HTML code of these webpages using functionality- and rendering-preserving manipulations so that they are classified as benign.

\myparagraph{Knowledge.} In our threat model, we assume a \emph{black-box} scenario \cite{Demontis:TransferabilityAdvAttacks, Biggio:Wild}.
Specifically, the machine-learning algorithm, its features, the parameters as well as the data, and the objective function used during the training phase are unknown to the attacker.
Regarding the feature set, although it is generally assumed that the attacker does not know the exact features used by the machine learning algorithm \cite{Biggio:Wild},
it is possible to obtain information about the most widely used features in the \sota by analyzing the description of many solutions that are publicly available in the literature (e.g., \cite{Apruzzese:SpacePhish, Sharma2020:FeatSelectionPhishing}).
Based on this idea, we have carefully analyzed the most common HTML features adopted in the literature and defined ad-hoc adversarial manipulations to evade all of them.
In this way, the attacker can use all the defined manipulations with the aim to evade as many features as possible.

\myparagraph{Capability.}
In our threat model, we assume that the attacker can use the ML-PWD as an \textit{oracle} by querying it and collecting its output confidence score, representing the probability of classifying the input webpages as phishing.

\myparagraph{Strategy.}
The adversarial phishing webpages can be generated by solving the following optimization problem:
\begin{mini}|l|
    {\vct t \in \mathcal{T}}{f ( h(\vct z, \vct t) )}{}{} \,,
    \label{eq:problem}
\end{mini}
which amounts to find the sequence of manipulations $\vct t = [t_0, \dots, t_K]$ that, when applied to the given phishing webpage $\vct z$, generate an adversarial phishing webpage, $z^{\star} = h(\vct z, \vct t)$, that minimizes the confidence score $f ( z^{\star} )$ returned by the target machine-learning model denoted with $f$.
For simplicity, in our formulation we assume that the machine-learning model includes a feature extraction step before classification, i.e., $f$ takes the raw webpage directly as input, but internally performs a preliminary step to map the input webpage to a feature vector.
Moreover, $h : \mathcal{Z} \times \mathcal{T} \rightarrow \mathcal{Z}$ is a function that applies a sequence of functionality- and rendering-preserving manipulations $\vct t$ to the HTML code of the phishing webpage $\vct z$, and outputs a valid webpage with the same behavior and rendering as the input one, but with a different HTML code.
Under the given black-box setting, and considering that the feature extraction step performed by $f$ may not be differentiable, the above optimization problem cannot be solved using classical gradient-based approaches.
For this reason, in this work we adopt a \emph{black-box (\aka gradient-free)} optimization algorithm that is described in detail in \autoref{sec:mutation-algo}.

\subsection{Comparison with \emph{SpacePhish}}
Our threat model differs from that proposed by Apruzzese \etal~\cite{Apruzzese:SpacePhish}, as we assume the possibility of querying the ML-PWD.
Recall indeed that this is a valid assumption adopted in many papers \cite{papernot2016transferability, bai2023query, cheng2019improving, liang2016cracking, li2014towards}, especially when considering Machine-Learning-as-a-Service (MLaaS) scenarios, in which the attacker can interact with the target machine-learning model by sending queries to it and observing its predictions~\cite{Oprea:AdvMLTaxonomyNIST, papernot2016transferability}.
For instance, those available through VirusTotal can be easily queried through dedicated APIs provided by the VirusTotal platform \cite{Peng:VTBlackBoxPhishing, Choo:VTPhishingStudy}.
In this work, we want to extend the threat model of \emph{SpacePhish} in order to thoroughly evaluate the adversarial robustness of \sota ML-PWD when the attacker can optimize the adversarial attacks.
Finally, it is worth noting that, even if the output of the ML-PWD is not available, the attacker can still optimize the adversarial attacks by using a so-called surrogate model~\cite{Demontis:TransferabilityAdvAttacks, Oprea:AdvMLTaxonomyNIST, demetrio2021functionality, Zhong:TransferableAdvAttacksDeepFace}. However, this approach is out of the scope of our work.
\section{Optimized HTML adversarial attacks}\label{sec:adv_attacks}
In this section, we describe our methodology for generating optimized and query-efficient HTML adversarial attacks.
Specifically, we first present our novel set of 14 functionality- and rendering-preserving adversarial manipulations designed to evade the HTML features described in \autoref{sec:html_features}.
Then, we describe our black-box optimizer that uses the proposed manipulations in order to optimize the generation of adversarial phishing webpages.

\subsection{Adversarial Manipulations}\label{sec:manipulations}
Each manipulation consists in a function that takes in input a phishing webpage, modifies its HTML code, and returns the new valid webpage with the same functionality and rendering as the input.
In the following we will describe the details of our manipulations, including the HTML features they aim to evade, as well as how they preserve the original rendering and functionality.

\begin{table}[!htpb]
  \centering
  \begin{tabular}{C{2.75cm} C{3.5cm} C{1.15cm}}
      \toprule
      \textbf{Manipulation}       &  \textbf{Evaded feature(s)}   &  \textbf{Type}  \\
      \midrule
      $\text{\emph{InjectIntElem}}^{\star}$         &  HTML\_freqDom, HTML\_objectRatio, HTML\_commPage, HTML\_nullLnkWeb \newline (int. links) &  MR  \\
      \midrule
      $\text{\emph{InjectIntElemFoot}}^{\star}$     &  HTML\_commPageFoot, HTML\_nullLnkFooter \newline (int. links)  &  MR  \\
      \midrule
      $\text{\emph{InjectIntLinkElem}}$     &  HTML\_metaScripts  &  MR  \\
      \midrule
      \emph{InjectExtElem}         &  HTML\_freqDom, HTML\_objectRatio, HTML\_metaScripts, HTML\_commPage &  MR  \\
      \midrule
      \emph{InjectExtElemFoot}     &  HTML\_commPageFoot &  MR  \\
      \midrule
      \emph{UpdateForm}           &  HTML\_SFH (int. links), HTML\_loginForm \newline (int. links) &  SR  \\
      \midrule
      \emph{ObfuscateExtLinks}     &  HTML\_SHF (ext. links), HTML\_brokenLnk, HTML\_anchors (ext. links), HTML\_css, \newline HTML\_favicon (ext. links), HTML\_loginForm \newline (ext. links)  &  SR  \\
      \midrule
      \emph{ObfuscateJS}          &  HTML\_statBar, HTML\_rightClick, HTML\_popUP  &  SR  \\
      \midrule
      \emph{InjectFakeCopyright}  &  HTML\_domCopyright  &  SR  \\
      \midrule
      \emph{UpdateIntAnchors}     &  HTML\_anchors (int. links), HTML\_nullLnkWeb (useless links), HTML\_nullLnkFooter (useless links)  &  SR  \\
      \midrule
      \emph{UpdateHiddenDivs}      &  HTML\_hiddenDiv  &  SR  \\
      \midrule
      \emph{UpdateHiddenButtons}   &  HTML\_hiddenButton  &  SR  \\
      \midrule
      \emph{UpdateHiddenInputs}    &  HTML\_hiddenInput  &  SR  \\
      \midrule
      \emph{UpdateTitle}          &  HTML\_URLBrand  &  SR  \\
      \midrule
      \emph{UpdateIFrames}         &  HTML\_iFrame  &  SR  \\
      \midrule
      \emph{InjectFakeFavicon}    &  HTML\_favicon \newline (no favicon included)  &  SR  \\
      \bottomrule
  \end{tabular}
\caption{Adversarial manipulations used in this work along with the corresponding evaded features and their type, defined according to the way they can be applied by the black-box optimizer (see \autoref{sec:mutation-algo}), \ie, single-round (SR) or multi-round (MR). The manipulations marked with $\star$ have been originally proposed by Apruzzese \etal \cite{Apruzzese:SpacePhish}.}
\label{tab:manipulations_evaded_features}
\end{table}

\myparagraph{\emph{InjectIntElem}.}
This manipulation aims to inject a given number of internal HTML elements into the body of the webpage.
It has been proposed in \emph{SpacePhish} to implement the $\text{WA}^r$ and $\widehat{\text{WA}^r}$ attacks with the aim to evade the HTML\_objectRatio feature \cite{Apruzzese:SpacePhish_Supp}.
The former, $\text{WA}^r$, assumes no knowledge about the target phishing detectors and injects 50 hidden anchors with internal links.
On the other hand, the latter, $\widehat{\text{WA}^r}$, assumes an attacker who knows how the HTML\_objectRatio feature works including its thresholds, hence this manipulation injects as many links as needed to meet the suspicious threshold ($0.15$) so that the sample is considered benign by this feature.
In our case, we also assume that the attacker does not know the internal thresholds used by the HTML\_objectRatio feature.
Therefore, in order to evade that feature, we design a black-box algorithm (see \autoref{sec:mutation-algo}) that iteratively applies this manipulation in order to inject a fixed number of internal elements, until the confidence score returned by the target phishing detector decreases, thus meaning that the feature has been evaded.
In our implementation, we inject the same type of HTML elements as in \emph{SpacePhish}, \ie, anchors, but the number of injected internal elements is set to 10 in order to have a finer level of granularity.
Using this manipulation, we are able to evade other HTML features that depend on anchor elements with internal links, \ie, HTML\_freqDom, HTML\_commPage, HTML\_nullLnkWeb.
Regarding the HTML\_nullLnkWeb feature, this manipulation only targets internal anchors included in the body or the footer.
On the other hand, to bypass this feature when it searches for patterns that represent useless internal links we have created another manipulation, \emph{UpdateIntAnchors}, which is described in the following.

Finally, since this manipulation injects some HTML elements, we must ensure that they are properly hidden in order to preserve the original rendering.
To this end, there are several approaches that can be adopted by the attacker (see \autoref{code:example_hiding}):
\begin{enumerate}
    \item Using the \lstinline[basicstyle=\ttfamily]{hidden} attribute (line \ref{code_line:hidden}). Inserting this attribute into an HTML element tells the browser to not render the content of the element. This is the default approach adopted by this manipulation.
    \item Modifying the style of the element. It is possible to hide an HTML element setting the \lstinline[basicstyle=\ttfamily]{style} attribute to "\lstinline[basicstyle=\ttfamily]{display:none}" (line \ref{code_line:style_inline}). This is the approach used in \emph{SpacePhish} \cite{Apruzzese:SpacePhish, Apruzzese:SpacePhish_Supp}.
    \item Similarly to (2), but using the \lstinline[language=html]{<style>} HTML element (lines \ref{code_line:style_elem_start}-\ref{code_line:style_elem_end}) instead of the \lstinline[basicstyle=\ttfamily]{style} attribute.
    \item Using \lstinline[language=html]{<noscript>} (lines \ref{code_line:noscript_start}-\ref{code_line:noscript_end}) and add inside it the HTML elements to be hidden. It is worth noting that this only works if \gls*{JS} is enabled on the victim's web browser.
\end{enumerate}

\myparagraph{\emph{InjectIntElemFoot}.}
This manipulation behaves similarly to \emph{InjectIntElem} but injects the internal elements into the footer of the webpage to evade the HTML\_commPageFoot and HTML\_nullLnkFooter features.

\myparagraph{\emph{InjectIntLinkElem}.}
This manipulation works exactly as \emph{InjectIntElem} but injects 10 hidden HTML elements of type \lstinline[language=html]{<link>} instead of \lstinline[language=html]{<a>} in order to evade the HTML\_metaScript feature since it depends on \lstinline[language=html]{<link>} elements.

\myparagraph{\emph{InjectExtElem}.}
This manipulation behaves similarly to \emph{InjectIntElem} but injects external HTML elements, \ie, elements with external links, instead of internal ones.
Specifically, it injects 10 \lstinline[language=html]{<link>} elements that are also hidden as for \emph{InjectIntElem} (\ie, by adding the \lstinline[basicstyle=\ttfamily]{hidden} attribute) to preserve the original rendering.
The injected external links are randomly extracted from a list of some well-known websites selected from the Alexa Top Million ranking\footnote{\url{https://www.alexa.com/}} in order to appear benign.
This manipulation evades multiple features that depend on external elements, which are HTML\_freqDom, HTML\_objectRatio, HTML\_commPage, and HTML\_metaScript.

\myparagraph{\emph{InjectExtElemFoot}.}
This manipulation works similarly as \emph{InjectExtElem}, but the external elements are inserted into the footer of the webpage with the goal to evade the HTML\_commPageFoot feature.

\myparagraph{\emph{UpdateForm}.}
This manipulation has been designed to evade the HTML\_SFH and HTML\_loginForm features when a form in the webpage includes an internal link matching one of the patterns searched by the two features, which represent useless internal links generally used by attackers such as \lstinline[basicstyle=\ttfamily]{#}.
Specifically, this manipulation replaces the original internal link, specified with \lstinline[basicstyle=\ttfamily]{action} attribute, with another random one that does not trigger the target features, such as \lstinline[basicstyle=\ttfamily]{#!} or \lstinline[basicstyle=\ttfamily]{#none}.
The original rendering is not affected because this manipulation updates a property of forms that does not affect the visual appearance of the webpage.

\myparagraph{\emph{ObfuscateExtLinks}.}
This manipulation aims to obfuscate the external links in a webpage in order to evade multiple HTML features, \ie, HTML\_SHF, HTML\_loginForm, HTML\_css, HTML\_anchors, HTML\_brokenLnk and HTML\_favicon.
Specifically, this manipulation executes the following steps:
\begin{enumerate}
    \item Substitute the external link with a random internal one that is not detected as suspicious by the HTML features (\lstinline[basicstyle=\ttfamily]{#!} as for HTML\_SHF);
    \item Create a new script element (\lstinline[language=html]{<script>}) that updates the value of the \lstinline[basicstyle=\ttfamily]{action} attribute to the original external link when the page is loaded;
    \item Add the new script element into the \lstinline[language=html]{<head>} of the webpage.
\end{enumerate}
To better explain the obfuscation approach, let's consider a practical example that shows how to evade the HTML\_SHF feature.
For instance, let's examine the simple webpage shown in \autoref{code:example_form}.
It includes a form (lines 7-10) with a malicious external link (line 7) for stealing the victim's credentials, which is detected by the HTML\_SHF feature.
Listing \ref{code:example_form_obfuscated} shows a new webpage in which the malicious link has been obfuscated using the script in lines 5-9.
In particular, the original link assigned to \lstinline{action} is updated with a random internal one (\lstinline[basicstyle=\ttfamily]{#!}),
but its original value is restored (line 7) when the page is loaded.
This new adversarial phishing webpage has the same rendering as the original one, but it is no longer detected by the HTML\_SHF.
Furthermore, this manipulation can be applied to obfuscate the external links included in any HTML elements, thus we use it to bypass multiple features as described in the following.
Regarding the HTML\_anchors feature, we use this manipulation to obfuscate the external links embedded in anchor elements, thus reducing the suspicious anchor rate computed by this feature.
In this way, the attacker is still able to insert hidden anchors with malicious external links but without being detected by the HTML\_anchors feature.
This manipulation can also evade the HTML\_brokenLnk feature by replacing all broken links (if any) with internal ones, hence resulting in a benign behavior for this feature.
The same applies to HTML\_loginForm, HTML\_css and HTML\_favicon, which can be evaded using this manipulation by obfuscating the external links analyzed by such features.
Finally, It is worth noting that, although this manipulation modifies external links, it is independent of \emph{InjectExtElem} and \emph{InjectExtElemFoot} because they target different features.
At the same time, the external links injected by these manipulations do not affect the features targeted by  \emph{ObfuscateExtLinks}.

\myparagraph{\emph{ObfuscateJS}.}
This manipulation aims to obfuscate the \acrfull*{JS} code inside the webpage inserted in \lstinline[language=html]{<script>} elements in order to evade the HTML\_popUP, HTML\_rightClick and HTML\_statBar features.
To achieve so, several techniques have been proposed in the literature~\cite{Bertholon2013JShadObfAJ, Wei:Javascript_Obfuscation}.
In this work, however, we use a different approach inspired to~\cite{Gao:MLSEC} for obfuscating the entire HTML code in a webpage, which is described in the following.
For instance, let us consider the webpage in Listing \ref{code:example_js}, which includes a script element to open a malicious webpage.
Because of the use of \lstinline[basicstyle=\ttfamily]{window.open()} DOM method, the webpage is considered malicious by the HTML\_popUP feature.
To bypass such feature, this manipulation operates as follows:
\begin{enumerate}
    \item Extracts the JS code from the original script and encodes it into Base64 \cite{base64_rfc}.
    \item Replaces the content of the original script with new JS code that creates a new script to hold the original JS code (line \ref{code_line:example_js_new_script}), decodes the original obfuscated JS code (line \ref{code_line:example_js_decode}), and insert the new script into the webpage to be executed (line \autoref{code_line:example_js_append}).
\end{enumerate}
It is worth noting that this approach can be also used to obfuscate the patterns searched by the other target features.
Moreover, the original rendering is preserved.

\myparagraph{\emph{InjectFakeCopyright}.}
This manipulation is used to evade the HTML\_domCopyright by injecting a new hidden paragraph containing the copyright symbol followed by the \lstinline[basicstyle=\ttfamily]{"Copyright"} string and the domain name of the website. 
For instance, assuming that the domain name of the webpage to manipulate is \lstinline[basicstyle=\ttfamily]{mydomain}, the injected element is: \lstinline[language=html, mathescape]{<p hidden>$\text{\copyright}$ Copyright mydomain </p>}.
Since the injected paragraph is hidden, the original rendering is preserved.

\myparagraph{\emph{UpdateIntAnchors}.}
This manipulation is designed to evade the HTML\_statBar, HTML\_nullLnkWeb and HTML\_nullLnkFooter features by replacing every useless internal link with another one that is not checked by such features, such as \lstinline[basicstyle=\ttfamily]{#!}.
The original rendering is preserved since this manipulation does not affect it by design.

\myparagraph{\emph{UpdateHiddenDivs}.}
This manipulation is designed to evade the HTML\_hiddenDiv feature by updating the way div elements (\lstinline[language=html]{<div>}) are hidden.
It operates in different ways according to how a div element is hidden, \ie, by setting the \lstinline[basicstyle=\ttfamily]{style} attribute to \lstinline[basicstyle=\ttfamily, breaklines]{visibility:hidden} or \lstinline[basicstyle=\ttfamily]{display:none}.
The main difference between the two approaches consists in how they allocate the space for the hidden element when rendering the webpage.
Specifically, the former (\ie, \lstinline[basicstyle=\ttfamily]{visibility:hidden}) still takes up space in the layout, while the latter (\ie, \lstinline[basicstyle=\ttfamily]{display:none}) does not take up any space.
For instance, let's consider Listing \ref{code:example_div} showing a div element hidden with \lstinline[basicstyle=\ttfamily]{display:none} (line \autoref{code_line:div_display_none}).
It can be removed and, to achieve the same behavior and rendering, we can insert the \lstinline[basicstyle=\ttfamily]{hidden} attribute (line \autoref{code_line:div_hidden} of Listing \ref{code:example_div_adv}) in order to evade the HTML\_hiddenDiv feature since it does not check for the presence of such attribute.
However, we cannot adopt the same approach for obfuscating div elements hidden using \lstinline[basicstyle=\ttfamily]{visibility:hidden} (line \autoref{code_line:div_vis_hidden} of Listing \ref{code:example_div}), because this will change the rendering.
In this case, we can still evade the HTML\_hiddenDiv feature by removing \lstinline[basicstyle=\ttfamily]{visibility:hidden} from the \lstinline{style} attribute and inserting a new \lstinline[language=html]{<style>} element to achieve the same result (lines \autoref{code_line:div_hidden_style_start}-\autoref{code_line:div_hidden_style_end} of Listing \ref{code:example_div_adv}).

\myparagraph{\emph{UpdateHiddenButtons}.}
This manipulation is designed to evade the HTML\_hiddenButton feature by obfuscating all the disabled button elements.
Specifically, for each disabled button, it removes the \lstinline[basicstyle=\ttfamily]{disabled} attribute and inserts a new script element that, by exploiting JS, adds this attribute back during rendering using the \lstinline[basicstyle=\ttfamily]{setAttribute()} DOM method.
Notably, this approach is similar to the one adopted by \emph{ObfuscateExtLinks} to obfuscate external links.
Thus, both the rendering and original behavior are preserved.

\myparagraph{\emph{UpdateHiddenInputs}.}
This manipulation consists of evading the HTML\_hiddenInput, and it operates in different ways according to whether the input element is hidden or disabled (since both are checked by the HTML\_hiddenInput feature).
Specifically, if the input element is hidden, this manipulation updates the value of its \lstinline{type} attribute from "hidden" to "text" and then adds the \lstinline{hidden} attribute.
Otherwise, if the input element is disabled, then this manipulation operates in the same way as \emph{UpdateHiddenButtons} by removing the attribute from the element and inserting it back during the rendering of the webpage by using \gls*{JS}.
In both cases, the original behavior and rendering remain the same.

\myparagraph{\emph{UpdateTitle}.}
This manipulation aims to evade the HTML\_URLBrand feature.
Specifically, if the website's domain name is not included in the title element, this manipulation updates the webpage title with the website's domain name and then replaces back the original title during rendering using a script element (\ie, similarly as how \emph{UpdateHiddenButtons} and \emph{UpdateHiddenInputs} work).

\myparagraph{\emph{UpdateIFrames}.}
This manipulation adopts the same approach of \emph{UpdateHiddenDivs}. Indeed, both the features look for the same patterns, but \emph{UpdateIFrames} targets \lstinline[language=html]{<iframe>} elements in order to evade the HTML\_iFrame feature.

\myparagraph{\emph{InjectFakeFavicon}.}
This manipulation is designed to inject a fake favicon in webpages that do not contain one, preventing them from being flagged as suspicious by the HTML\_favicon feature.
Specifically, this manipulation injects a favicon element with a useless internal link, such as \ie, \lstinline[language=html]{<link rel="icon" href="#none">}, into the head of the webpage.

\begin{lstlisting}[
   language=HTML,
   float=h,
   basicstyle=\fontsize{7.5}{8.5}\selectfont\ttfamily,
   commentstyle=\color{gray},
   showspaces=false,
   showstringspaces=false,
   breakindent=1em,
   breaklines=true,
   numbers=left,
   xleftmargin=1.6em,
   keepspaces=true,
   frame=single,
   caption={Example showing different approaches to hide HTML elements: using the \inlinecode{hidden} attribute (line 10), modifying the CSS style (lines 6 and 11), and embedding the element in \inlinehtml{<noscript>} (line 14).},
   captionpos=b,
   escapechar=|,
   label={code:example_hiding}
]
<!DOCTYPE html>
<html>
<head>
<title>Home</title>
<style>    |\label{code_line:style_elem_start}|
  #mypar {display: none;}
</style>    |\label{code_line:style_elem_end}|
</head>
<body>
  <p hidden="">Hidden text</p>  |\label{code_line:hidden}|
  <p style="display:none">Hidden text</p>  |\label{code_line:style_inline}|
  <p id="mypar">Hidden text</p>  |\label{code_line:style}|
  <noscript>  |\label{code_line:noscript_start}|
    <p>Hidden text</p>
  </noscript>    |\label{code_line:noscript_end}|
</body>
</html>
\end{lstlisting}

\begin{lstlisting}[
   language=HTML,
   float=h,
   basicstyle=\fontsize{7.5}{8.5}\selectfont\ttfamily,
   commentstyle=\color{gray},
   showspaces=false,
   showstringspaces=false,
   breakindent=1em,
   breaklines=true,
   numbers=left,
   xleftmargin=1.6em,
   keepspaces=true,
   frame=single,
   escapechar=|,
   caption={Webpage including a form with a malicious external link (line 7) detected by the HTML\_SHF feature.},
   captionpos=b,
   label={code:example_form}
]
<!DOCTYPE html>
<html>
<head>
<title>Login</title>
</head>
<body>
  <form id="myform" action="http://malicious.io">
    <label for="pwd">Enter your password: </label>
    <input type="password" name="pass" required>
  </form>
</body>
</html>
\end{lstlisting}

\begin{lstlisting}[
   language=HTML,
   float=h,
   basicstyle=\fontsize{7.5}{8.5}\selectfont\ttfamily,
   commentstyle=\color{gray},
   showspaces=false,
   showstringspaces=false,
   breakindent=1em,
   breaklines=true,
   numbers=left,
   xleftmargin=1.6em,
   keepspaces=true,
   frame=single,
   caption={Adversarial phishing webpage generated using \emph{ObfuscateExtLinks}, which obfuscates the malicious link (lines 5 - 9) in the original webpage of Listing \ref{code:example_form}.},
   captionpos=b,
   escapechar=|,
   label={code:example_form_obfuscated}
]
<!DOCTYPE html>
<html>
<head>
<title>Login</title>
<script type="text/javascript">
window.onload = function () {
    document.getElementById("myform").setAttribute("action", "http://malicious.io");
}
</script>
</head>
<body>
  <form id="myform" action="#!">
    <label for="pwd">Enter your password: </label>
    <input type="password" name="passwd" required>
  </form>
</body>
</html>
\end{lstlisting}

\begin{lstlisting}[
   language=HTML,
   float=h,
   basicstyle=\fontsize{7.6}{8.6}\selectfont\ttfamily,
   commentstyle=\color{gray},
   showspaces=false,
   showstringspaces=false,
   breakindent=1em,
   breaklines=true,
   numbers=left,
   xleftmargin=1.6em,
   keepspaces=true,
   frame=single,
   caption={Webpage using \emph{window.open()} to load an external malicious link (line 5) detected by the HTML\_popUP feature.},
   captionpos=b,
   escapechar=|,
   label={code:example_js}
]
<html>
<head>
<title>Home</title>
<script>
  window.open("http://malicious.io", "_self");
</script>
</head>
<body>
</body>
</html>
\end{lstlisting}

\begin{lstlisting}[
   language=HTML,
   float=h,
   basicstyle=\fontsize{7.6}{8.6}\selectfont\ttfamily,
   commentstyle=\color{gray},
   showspaces=false,
   showstringspaces=false,
   breakindent=1em,
   breaklines=true,
   numbers=left,
   xleftmargin=1.6em,
   keepspaces=true,
   frame=single,
   caption={Adversarial phishing webpage manipulated using \emph{ObfuscateJS} in order to obfuscate the JS code (lines 4 - 9) of the webpage shown in Listing \ref{code:example_js}.},
   captionpos=b,
   escapechar=|,
   label={code:example_js_adv}
]
<html>
<head>
<title>Home</title>
<script>
  let script = document.createElement("script");  |\label{code_line:example_js_new_script}|
  script.innerHTML = atob("d2luZG93Lm9wZW4oImh0 \ |\label{code_line:example_js_decode}|
    dHA6Ly9tYWxpY2lvdXMuaW8iLCAiX3NlbGYiKTs=");
  document.head.append(script);  |\label{code_line:example_js_append}|
</script>
</head>
<body>
</body>
</html>
\end{lstlisting}

\begin{lstlisting}[
   language=HTML,
   float=h,
   basicstyle=\fontsize{7.6}{8.6}\selectfont\ttfamily,
   commentstyle=\color{gray},
   showspaces=false,
   showstringspaces=false,
   breakindent=1em,
   breaklines=true,
   numbers=left,
   xleftmargin=1.6em,
   keepspaces=true,
   frame=single,
   caption={Webpage with two hidden div HTML elements (lines 7 and 11) detected by the HTML\_hiddenDiv feature.},
   captionpos=b,
   escapechar=|,
   label={code:example_div}
]
<!DOCTYPE html>
<html>
<head>
  <title>Home</title>
</head>
<body>
  <div id="div1" style="display: none">  |\label{code_line:div_display_none}|
    <p>Text in the first div.</p>
  </div>

  <div id="div2" style="visibility: hidden">  |\label{code_line:div_vis_hidden}|
    <p>Text in the second div.</p>
  </div>
</body>
</html>
\end{lstlisting}

\begin{lstlisting}[
   language=HTML,
   float=h,
   basicstyle=\fontsize{7.6}{8.6}\selectfont\ttfamily,
   commentstyle=\color{gray},
   showspaces=false,
   showstringspaces=false,
   breakindent=1em,
   breaklines=true,
   numbers=left,
   xleftmargin=1.6em,
   keepspaces=true,
   frame=single,
   escapechar=|,
   caption={Adversarial webpage generated by manipulating the webpage of Listing \ref{code:example_div} through \emph{UpdateHiddenDivs}, which hides the div elements using CSS combined with the \inlinehtml{<style>} element (line 6), and the \inlinecode{hidden} attribute (line 10).},
   captionpos=b,
   label={code:example_div_adv}
]
<!DOCTYPE html>
<html>
<head>
<title>Home</title>
<style>   |\label{code_line:div_hidden_style_start}|
  #div2 {visibility: hidden;}
</style>  |\label{code_line:div_hidden_style_end}|
</head>
<body>
  <div id="div1" hidden>  |\label{code_line:div_hidden}|
    <p>Text in the first div.</p>
  </div>

  <div id="div2">
    <p>Text in the second div.</p>
  </div>
</body>
</html>
\end{lstlisting}

\subsection{Mutation-based Black-box Optimizer}\label{sec:mutation-algo}
To optimize the choice of the manipulations defined in \autoref{sec:manipulations}, we propose a black-box optimizer (shown in Algorithm \ref{algo:bb_improved}) that is in line with the proposed threat model (see \autoref{sec:threat_model}).
Our optimizer draws inspiration from the algorithm proposed in \emph{\wafamole} \cite{Demetrio:Wafamole2020}, which relies on mutation-based fuzzing techniques~\cite{fuzzingbook2023:MutationFuzzer}, recently shown to be promising for generating adversarial examples~\cite{Park:GradFuzz2023, Demetrio:Wafamole2020}.
Specifically, the algorithm of \emph{\wafamole} adopts an iterative approach consisting of consecutive mutation rounds with the aim to mutate the original malicious sample in order to minimize the confidence score returned by the machine-learning model.
Starting from the original algorithm of \emph{\wafamole}, we have designed a novel one that is tailored to the proposed manipulations in order to improve its effectiveness, \ie, minimize the number of queries when generating the adversarial attacks.
To this end, in the following, we first explain how the manipulations can be categorized in order to make the optimizer more query-efficient, and then we describe how the optimizer works step-by-step.

\myparagraph{Categorization of the HTML Manipulations.}
According to how the proposed manipulations can be applied to the input phishing webpage, they can be categorized into two main classes: single-round (SR), if they can be applied for just a single mutation round, or multi-rounds (MR), if they require more sequential mutation rounds.
Specifically, SR manipulations generate the same output (\ie, a manipulated webpage) when used sequentially for more than one round, so it is sufficient to use them for a single round.
On the other hand, this does not apply to MR manipulations, whose output can change at each round.
Furthermore, SR manipulations are independent of each other, while MR manipulations can be correlated, \ie, they can impact a common set of features.

To better explain the difference between the two classes, let's consider some of the manipulations defined in \autoref{sec:manipulations}.
For instance, the \emph{UpdateHiddenDivs} is an SR manipulation because, after it is used for the first time, all the related div elements are updated and there is no need to use it in the next rounds since no other manipulation can inject hidden div elements that may trigger the features (\ie, HTML\_hiddenDiv) targeted by this manipulation.
The same applies to other manipulations such as \emph{UpdateHiddenButtons}, \emph{UpdateHiddenInputs} and \emph{UpdateTitle}.
On the contrary, manipulations like \emph{InjectIntElem} and \emph{InjectExtElem} belong to the MR class because, in general, they need to be applied in multiple consecutive rounds to effectively evade the target HTML features.
For instance, let's consider the HTML\_commPage.
In order to evade this feature, the attacker has to apply both \emph{InjectIntElem} and \emph{InjectExtElem} for multiple consecutive rounds to find the proper ratio between internal and external links.
Clustering manipulations into the two defined classes offers a significant advantage in enhancing the optimizer's efficiency.
Indeed, if using the approach used in WAF-a-MoLE, which randomly selects manipulations for each mutation round, there's a risk of applying the same SR manipulation repeatedly in consecutive rounds, resulting in a significant waste of queries because the webpage would not be updated.
Conversely, to address this issue our optimizer first executes the SR manipulations one by one, and then runs the main loop of mutational rounds by using only the MR manipulations.

\myparagraph{Algorithm Description.}
Initially, the optimizer initializes the best adversarial example $\vct z^{\star}$ and score $s^{\star}$ found so far with the initial phishing webpage $\vct z$ (line \ref{line:bb_opt_init_z}), and its score $f ( \vct z^{\star} )$ (line \ref{line:bb_opt_init_score}).
Then, it sequentially applies the SR manipulations (lines \ref{line:bb_opt_sr_loop}-\ref{line:bb_opt_sr_classification}) and updates the best adversarial example and score found so far each time it finds a new manipulation that reduces the best score found so far (lines \ref{line:bb_opt_sr_eval}-\ref{line:bb_opt_sr_update_adv}).
Then, the optimizer executes the loop related to MR manipulations, which consists of $R$ mutation rounds (line \ref{line:bb_opt_loop}).
Specifically, during each mutation round, the algorithm generates new candidates (\ie, adversarial phishing webpages) from the current best adversarial example by using one MR manipulation for each candidate (lines \ref{line:bb_opt_mr_loop}-\ref{line:bb_opt_mr_add_candidate}).
Afterward, the algorithm selects the candidate having the lowest confidence score (line \ref{line:bb_opt_mr_find_best}) and, in case its score is lower than the best score found so far (line \ref{line:bb_opt_mr_check_score}), the chosen becomes the best adversarial example found so far (line \ref{line:bb_opt_mr_update_adv}).
Finally, regarding the choice of the number of mutation rounds $R$, given the maximum query budget $Q$,
it can be set using the following formula: $R = (Q - \#SR) \, / \, \#MR$, where $\#SR$ and $\#MR$ are the number of SR and MR manipulations, respectively.

\begin{algorithm}[!htb]
  \SetKwInOut{Input}{Input}
  \SetKwInOut{Output}{Output}
  \KwData{
  $\vct z$, the initial phishing sample; \\
  \hspace{\parindent} $f$, the machine-learning phishing webpage detector; \\
  \hspace{\parindent} $h$, the function to mutate the phishing webpages; \\
  \hspace{\parindent} $R$, the number of mutation rounds; \\
  \hspace{\parindent} $SR$ the set of single-round (SR) manipulations; \\
  \hspace{\parindent} $MR$ the set of multi-round (MR) manipulations.
  }
  \KwResult{$\vct z^\star$, the adversarial phishing sample.}
  $\vct z^{\star} = \vct z$ \label{line:bb_opt_init_z} \\
  $\text{s}^{\star} = f ( \vct z^{\star} )$  \label{line:bb_opt_init_score} \\

  \textbf{for} $ t $ \textbf{in} $ SR $ \label{line:bb_opt_sr_loop} \\
  \Indp
    $\vct z^{\prime} = h(\vct z^{\star}, [t])$ \label{line:bb_opt_sr_mutation} \\
    $\text{s}^{\prime} = f ( \vct z^{\prime} )$ \label{line:bb_opt_sr_classification} \\
    \textbf{if} $ \text{s}^{\prime} < \text{s}^{\star} $ \label{line:bb_opt_sr_eval} \\
    \Indp
        $\text{s}^{\star} = \text{s}^{\prime}$ \label{line:bb_opt_sr_update_score} \\
        $\vct z^{\star} = \vct z^{\prime}$ \label{line:bb_opt_sr_update_adv} \\
    \Indm
  \Indm

  \textbf{for} $ r $ \textbf{in} $ [1, \, R] $ \label{line:bb_opt_loop} \\
  \Indp
    $C = \emptyset$ \\
    \textbf{for} $ t $ \textbf{in} $ MR $ \label{line:bb_opt_mr_loop} \\
    \Indp
      $\vct z^{\prime} = h(\vct z^{\star}, [t])$ \label{line:bb_opt_mr_mutation} \\
      $\text{s}^{\prime} = f ( \vct z^{\prime} )$ \label{line:bb_opt_mr_classification} \\
      $C = C \cup \{ (\vct z^{\prime}, \text{s}^{\prime}) \}$ \label{line:bb_opt_mr_add_candidate} \\
    \Indm

    $\vct z^{b}, \, \text{s}^{b} = get\_best\_candidate(C)$ \label{line:bb_opt_mr_find_best} \\
    \textbf{if} $ \text{s}^{b} < \text{s}^{\star} $ \label{line:bb_opt_mr_check_score} \\
    \Indp
      $\text{s}^{\star} = \text{s}^{b}$ \label{line:bb_opt_mr_update_score} \\
      $\vct z^{\star} = \vct z^{b}$ \label{line:bb_opt_mr_update_adv} \\
    \Indm
  \Indm

  \textbf{return} $\vct z^\star$
  \caption{Mutation-based black-box optimizer to generate adversarial phishing webpages.}
  \label{algo:bb_improved}
\end{algorithm}
\section{Experimental analysis}\label{sec:experiments}
In this section, we first describe the setup adopted in our experiments, and then we present and discuss the obtained results.

\subsection{Experimental Setup}\label{sec:exp_setup}
We now present the setup underlying our experimental analysis, conducted on an Ubuntu 18.04.6 LTS server equipped with an Intel Xeon E7-8880 CPU (16 cores) and 64 GB of RAM.

\myparagraph{ML Algorithms.}
We evaluate the same machine-learning algorithms used in \emph{SpacePhish} \cite{Apruzzese:SpacePhish}:
\begin{itemize}
    \item Logistic Regression ($LR$), a linear model also adopted in the Google phishing page filter \cite{liang2016cracking, song2021advanced};
    \item Random Forest ($RF$), a tree-based ensemble learning algorithm \cite{breiman2001rf} that has been shown outstanding performance in phishing detection tasks \cite{Tian:TrackingElitePhishing};
    \item Convolutional Neural Network ($CNN$), a deep learning \cite{deeplearningbook} model used in \cite{Wei:CnnUrlPhishing} for detecting phishing webpages.
\end{itemize}
As for the feature set, we train each algorithm on the HTML features as well as the combination of both HTML and URL features, which are identified in \emph{SpacePhish} as $F^r$ and $F^c$, respectively \cite{Apruzzese:SpacePhish}.
The main reason for this choice is to assess the effectiveness of our adversarial attacks, particularly when incorporating supplementary features beyond those derived from the HTML code.

\myparagraph{Dataset.}
We evaluate our approach on the \emph{DeltaPhish} dataset~\cite{Corona:Deltaphish}, consisting of 5511 benign and 1012 phishing webpages.
We perform a stratified random split (to preserve the original ratio between benign and phishing distributions) by using the 80:20 ratio, which is commonly used in related literature \cite{Bac:PWDGAN, AlQurashi:OptimalAttackPaths2021}.
In other words, 80\% of both benign and phishing samples are used to build the training set, while the remaining 20\% of samples are part of the test set.

\myparagraph{Generation of Adversarial Phishing Webpages.}
We adopt the same approach of Apruzzese \etal~\cite{Apruzzese:SpacePhish}.
In particular, we randomly select from the test set 100 phishing samples that are correctly classified by the best ML-PWD (typically $F^c$).
Such 100 samples are used to evaluate the baseline detection rate of the target ML-PWD (\ie, \texttt{no-atk}), as well as to craft the adversarial examples using both the HTML adversarial attacks proposed in this work (\texttt{our}) and in \emph{SpacePhish} (\ie, $\text{WA}^r$ and $\widehat{\text{WA}^r}$) \cite{Apruzzese:SpacePhish}.
We would like to remind the reader that $\text{WA}^r$ consists of injecting 50 hidden internal links, while $\widehat{\text{WA}^r}$ injects as many internal links as needed to meet the suspicious threshold ($0.15$) of the HTML\_objectRatio feature.
As for our approach, the query budget for optimizing the adversarial attacks is set to $36$ queries, which implies $5$ mutation rounds (\ie, $R = 5$ in Algorithm \ref{algo:bb_improved}).

\subsection{Results and Discussion}\label{sec:results}
\begin{figure}[h]
    \centering
    \includegraphics[width=\columnwidth]{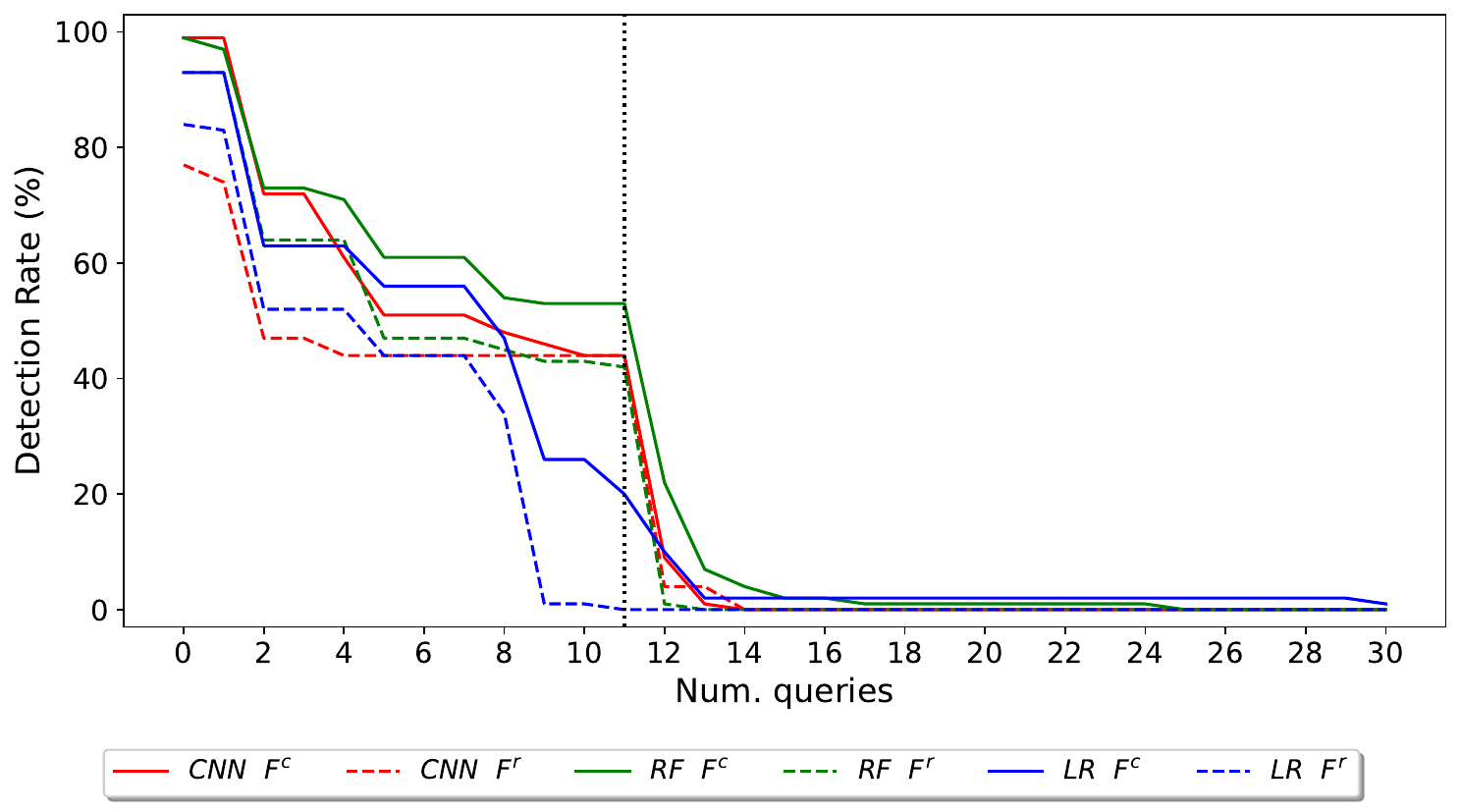}
    \caption{Security evaluation curves showing how the detection rate at 1\% FPR of the target ML-PWD changes \wrt the number of queries when applying the best sequence of manipulations.
    Flat regions in the plot indicate manipulations that are not applied because they do not decrease the output score.
    The impact of SR and MR manipulations is shown on the left and right sides of the dotted vertical line, respectively.}
    \label{fig:sec_eval_curves}
\end{figure}

The experimental results are reported in \autoref{tab:perf_comparison} and \autoref{fig:sec_eval_curves}.
The former shows the detection rate of the evaluated ML-PWD ($CNN$, $RF$ and $LR$) on the baseline test set of 100 samples (\texttt{no-atk}), as well as their adversarial robustness against the attacks proposed in \emph{SpacePhish} ($\text{WA}^r$ and $\widehat{\text{WA}^r}$) in this work (\texttt{our}).
The latter, instead, reports the security evaluation curves that show the detection rate at $1\%$ \gls*{FPR} of the target ML-PWD \wrt the number of queries when the best sequence of manipulations is applied.
It is worth noting that the drops in the detection rate represent manipulations that are effective in decreasing the confidence score and thus are included in the best (\ie, optimal) sequence of manipulations.
Instead, flat regions indicate manipulations that are ineffective and thus are not used to generate the final adversarial example.
Moreover, we have computed the detection rate at $1\%$ FPR because this threshold is widely adopted in the literature \cite{Corona:Deltaphish, Demontis2017YesML} as well as to perform a fair evaluation of the ML-PWD, \ie, they are evaluated assuming the same FPR.
From the obtained results we can gain several takeaways that are described in the following.

\myparagraph{Query-efficient Adversarial Attacks.}
The obtained results highlight that the proposed adversarial attacks clearly \emph{raze to the ground} the detection rate of all the evaluated ML-PWD using just 30 queries, hence underlining the effectiveness of the proposed methodology.
Specifically, by only using the SR manipulations (\ie, the first 11 queries shown on the left of the dotted vertical line in \autoref{fig:sec_eval_curves}) the average detection rate is lower than 50\% for all the ML-PWD except for the $RF$ model trained on the whole set of feature ($F^c$), whose detection rate is 53\%.
As for the MR manipulations, they play a crucial role in boosting the attack's effectiveness.
Indeed, as depicted in \autoref{fig:sec_eval_curves}, finding the optimal number of internal and external elements to inject significantly reduces the detection rate to nearly zero within just a few queries.
This also underlines that the HTML features related to the number of internal and external elements play a critical role in terms of adversarial robustness.

\myparagraph{HTML Features Matter.}
Even more interesting is the fact that the proposed adversarial manipulations, while targeting the HTML features, have proven effective in evading the ML-PWD trained on the whole feature set $F^c$, including both the HTML and URL features.
This underlines two key points. First, the adversarial robustness mainly relies on the HTML features, as also discussed above when analyzing the manipulations' effectiveness. Second, the supplementary URL features do not provide substantial benefits in terms of adversarial robustness.
Indeed, an attacker can effectively evade the ML-PWD by exclusively leveraging the proposed manipulations targeting the HTML features.

\begin{table}[h]
    \centering
    \resizebox{0.85\columnwidth}{!}{
        \begin{tabular}{c|c?cccc}
            \toprule
            $\text{ML algo}$ & $F$ & \texttt{no-atk} & $\text{WA}^r$ & $\widehat{\text{WA}^r}$ & \texttt{our} \\
            \addlinespace[0.2em]
            \toprule

            \multirow{2}{*}{{\smamath{CNN}}}
             & $F^r$ & $0.81$ & $0.33$ & $0.78$ & $\mathbf{0.00}$ \\
             & $F^c$ & $0.94$ & $0.93$ & $0.90$ & $\mathbf{0.00}$ \\
            
            \midrule 
            
            \multirow{2}{*}{{\smamath{RF}}}
             & $F^r$ & $0.95$ & $0.90$ & $0.79$ & $\mathbf{0.00}$ \\
             & $F^c$ & $0.97$ & $0.96$ & $0.90$ & $\mathbf{0.00}$ \\
            
            \midrule
            
            \multirow{2}{*}{{\smamath{LR}}}
             & $F^r$ & $0.72$ & $0.51$ & $0.53$ & $\mathbf{0.00}$ \\
             & $F^c$ & $0.86$ & $0.77$ & $0.72$ & $\mathbf{0.00}$ \\
            
            \bottomrule
        \end{tabular}
    }
    \caption{Average detection rate at 1\% FPR of the target ML-PWD ($CNN$, $RF$ and $LR$) on the \emph{DeltaPhish} dataset.
    Columns represent the baseline (\texttt{no-atk}), the attacks proposed in \emph{SpacePhish} ($\text{WA}^r$ and $\widehat{\text{WA}^r}$) \cite{Apruzzese:SpacePhish}, and our approach (\texttt{our}). \\
    The best results are in bold.}
    \label{tab:perf_comparison}
\end{table}

\section{Conclusions and Future Work}\label{sec:conclusions}
In this work, we have introduced a novel methodology for generating query-efficient and notably effective HTML adversarial attacks.
Specifically, we have designed a novel set of 14 functionality- and rendering-preserving manipulations that extend the current \sota, as well as a novel black-box optimizer tailored to such manipulations in order to generate adversarial phishing webpages that are able to \emph{raze to the ground} several \sota machine-learning phishing webpage detectors (ML-PWD).
Our experiments also reveal that the ML-PWD's adversarial robustness primarily depends on the HTML features as our methodology effectively evades detection even when using additional URL features.
To counter the adversarial attacks proposed in this work, a future work development is experimenting with well-known \sota approaches for increasing the adversarial robustness such as adversarial training \cite{madry2018advtrain, zheng2020efficientadvtrain} and certified robustness techniques~\cite{mueller2023certified}.
Moreover, although the HTML manipulations are specifically crafted to evade the features used in \emph{SpacePhish} \cite{Apruzzese:SpacePhish_Supp}, another interesting future work is evaluating our methodology \emph{in the wild}, \ie, assessing its effectiveness against production-grade phishing detectors, as well as other feature representations proposed in the literature.
Finally, as for the proposed black-box optimizer, while it leverages the output scores to optimize the selection of the adversarial manipulations, in principle, it can be also extended to the \emph{hard-label} scenario \cite{Satya:HardLableBBAttacks}.

\begin{acks}
This research has been supported by the TESTABLE project, funded by the European Union's Horizon 2020 research and innovation program (grant no. 101019206);
by Fondazione di Sardegna under the project ``TrustML: Towards Machine Learning that Humans Can Trust’’, CUP: F73C22001320007; and 
by project SERICS (PE00000014) under the MUR National Recovery and Resilience Plan funded by the European Union – NextGenerationEU.
\end{acks}

\onecolumn
\begin{multicols}{2}
\bibliographystyle{ACM-Reference-Format}
\bibliography{bibliography}
\end{multicols}

\end{document}
\endinput